\documentclass[a4paper,letters,fleqn,usenatbib]{mnras}
\usepackage[T1]{fontenc}
\usepackage{ae,aecompl}
\usepackage{enumitem}

\newcommand{\lsp}{LS~I~+61$^{\circ}$303}
\newcommand{\lsi}{LS~I~+61$^{\circ}$303~}

\newcommand{\grs}{GRS~1915+105~}

\newcommand{\beq}{\begin{equation}}
\newcommand{\eneq}{\end{equation}}
\newcommand{\fermilat}{\textit{Fermi}-LAT~}
\newcommand{\swift}{\textit{Swift}-XRT~}
\newcommand{\swiftp}{\textit{Swift}-XRT}
\newcommand{\nus}{\textit{NuSTAR}~}
\newcommand{\xmm}{\textit{XMM-Newton}~}
\usepackage{graphicx}	
\usepackage{amsmath}	
\usepackage{amssymb}	
\usepackage{xcolor}
\title{Evidence for Periodic Accretion-Ejection in \lsp}

\author[M. Massi et al.] {
M. Massi,$^{1}$\thanks{E-mail: mmassi@mpifr-bonn.mpg.de} M. Chernyakova,$^{2,3}$ A. Kraus,$^{1}$  D. Malyshev,$^{4}$ F. Jaron,$^{1}$ 
\and
S. Kiehlmann,$^{5,6,7}$  
~S. A. Dzib,$^{1}$   R. Sharma,$^{1}$  S. Migliari$^{8,9}$ and A. C. S. Readhead$^{7}$
\\
$^{1}$Max-Planck-Institut f\"ur Radioastronomie, Auf dem H\"ugel 69, D-53121 Bonn, Germany.\\
$^{2}$School of Physical Sciences and CfAR, Dublin City University, Dublin 9, Ireland\\
$^{3}$Dublin Institute for Advanced Studies, 31 Fitzwilliam Place, Dublin 2, Ireland.\\
$^{4}$Institut f\"ur Astronomie und Astrophysik, Universit\"at T\"ubingen,  D-72076 Germany.\\
$^{5}$Institute of Astrophysics, Foundation for Research and Technology-Hellas, GR-71110 Heraklion, Greece.\\
$^{6}$Department of Physics, Univ. of Crete, GR-70013 Heraklion, Greece.\\
$^{7}$OVRO, California Inst. of Technology, 1216 E California Blvd, Pasadena, CA 91125, USA.\\
$^{8}$ESAC/ESA, Camino Bajo del Castillo s/n, Urb. Villafranca del Castillo, 28692 Madrid, Spain.\\
$^{9}$Institute of Cosmos Sciences, Univ. of Barcelona, Mart\'{\i} i Franqu\`es 1, 08028 Barcelona, Spain.\\
}
\date{Accepted XXX. Received YYY; in original form ZZZ}

\pubyear{2019}

\begin{document}

\label{firstpage}
\pagerange{\pageref{firstpage}--\pageref{lastpage}}
\maketitle

\begin{abstract}
The stellar binary system \lsp, composed of a compact object in an 
eccentric orbit around a B0 Ve star,  emits  from radio up to $\gamma$-ray energies.
The orbital modulation of radio spectral index, X-ray, and GeV $\gamma$-ray data 
 suggests the presence of  two peaks. This two-peaked profile is 
in line with the accretion theory predicting two accretion-ejection
events for \lsi along the 26.5 days orbit.
However, the existing multiwavelength data are not simultaneous. 
In this paper we report the results of a campaign
covering radio, X-ray, and $\gamma$-ray observations of the system along one single orbit.
Our results   confirm the two predicted events along the orbit and
in addition show that the positions of radio and $\gamma$-ray peaks
 are coincident with X-ray dips
as expected for radio and $\gamma$-ray emitting ejections
depleting the X-ray emitting accretion flow.
We discuss 
future observing strategies for a systematic study of the
 accretion-ejection physical processes in \lsp.
\end{abstract}


\section{Introduction}

The stellar binary system \lsi consists of a compact object in an eccentric orbit ($e \simeq 0.7$) around a B0 Ve star \citep{casares05} with orbital period $P_{\rm orbit} = 26.496 \pm 0.0028$\,days\ \citep{gregory02}. It has been proposed that the compact object in the system could be a black hole \citep[e.g.,][]{punsly99} or a neutron star \citep[e.g.,][]{maraschitreves81}. On the other hand, the correlation between the X-ray luminosity and the X-ray spectral slope in \lsi agrees with that of black holes \citep{massiet17, massi17}.
\citet{dubus06} investigated if, in analogy to the pulsar binary PSR~B1259-63, 
the relativistic wind of a fast rotating young pulsar interacting with the Be wind 
could  explain the radio emission in \lsp. 
However, the radio characteristics of 
 the  periodic ($\sim P_{\rm orbit}$) outburst in \lsi  
 are different from the simple optically thin outburst of PSR~B1259-63 \citep{connors02} but
are consistent with a microquasar scenario, i.e., the scenario of  accreting objects  with associated jets
established for several black holes.
The outburst in \lsi has in fact the same complex structure as outbursts 
of the microquasars  XTE~J1752-223 and Cygnus~X-3 \citep{zimmermann15}:
After a flat or inverted outburst (radio spectral index $\alpha \geq 0$, i.e.\ same flux density for all frequencies or dominating at higher frequencies), there is an optically thin outburst (i.e.\ $\alpha < 0$, low frequencies dominating the higher ones, \citealt{massikaufman09,massi11}).
 
\citet{dubus06} analysed how the shocked material in the pulsar scenario expands all along the orbit, creating a radio nebula with a ``comma shape'' reminiscent of a one-sided jet. In particular, Dubus performed simulations for the system\ \lsi and predicts, at a scale larger than the orbital size, that the nebula should look like a one-sided jet with very stable position angle, reflecting the projection of the major axis of the orbit. 
Same conclusions are given in  \citet{moldon12},
modelling the flow of electrons accelerated in a pulsar wind/stellar wind interaction.
They find that the orientation of the resulting structures,
that  extend  to size scales much larger than the orbital size,  depend in fact on the orbital inclination.
Several VLBI observations of \lsi are available, all of them with  a beam 
 larger than the orbital size. The VLBI images show a radio jet with fast variations: A MERLIN map of \lsi shows a two-sided structure, with an S-shape as the precessing jet of SS433 \citep{hjellming88}, one day later a second MERLIN map shows a significantly rotated 
one-sided structure \citep{massietal04}. Also VLBA observations show changes from one-sided to double-sided jet, compatible with variable Doppler boosting due to changes of the jet orientation with respect to the line of sight \citep{massi12}. VLBA astrometry indicates that the variations are periodic: the jet core traces a closed path with  a period of $P_{\rm precession} = 26.926 \pm 0.005$\,days \citep{wu18}. 
Simulations of the young pulsar model predict  a stability in the orientation of the nebula 
 contradicted by the obervations of \lsp,  whereas the microquasar model  explains the variations 
with a precessing jet \citep[e.g., see Fig. 7 in][]{massitorricelli14}.
SS433  \citep{rasul19} and \lsi  
\citep[]{jaron14,jaron16,jaron18} are the two microquasars  
where jet precession can be traced up to the GeV band.

\begin{figure}
\includegraphics[scale=0.96]{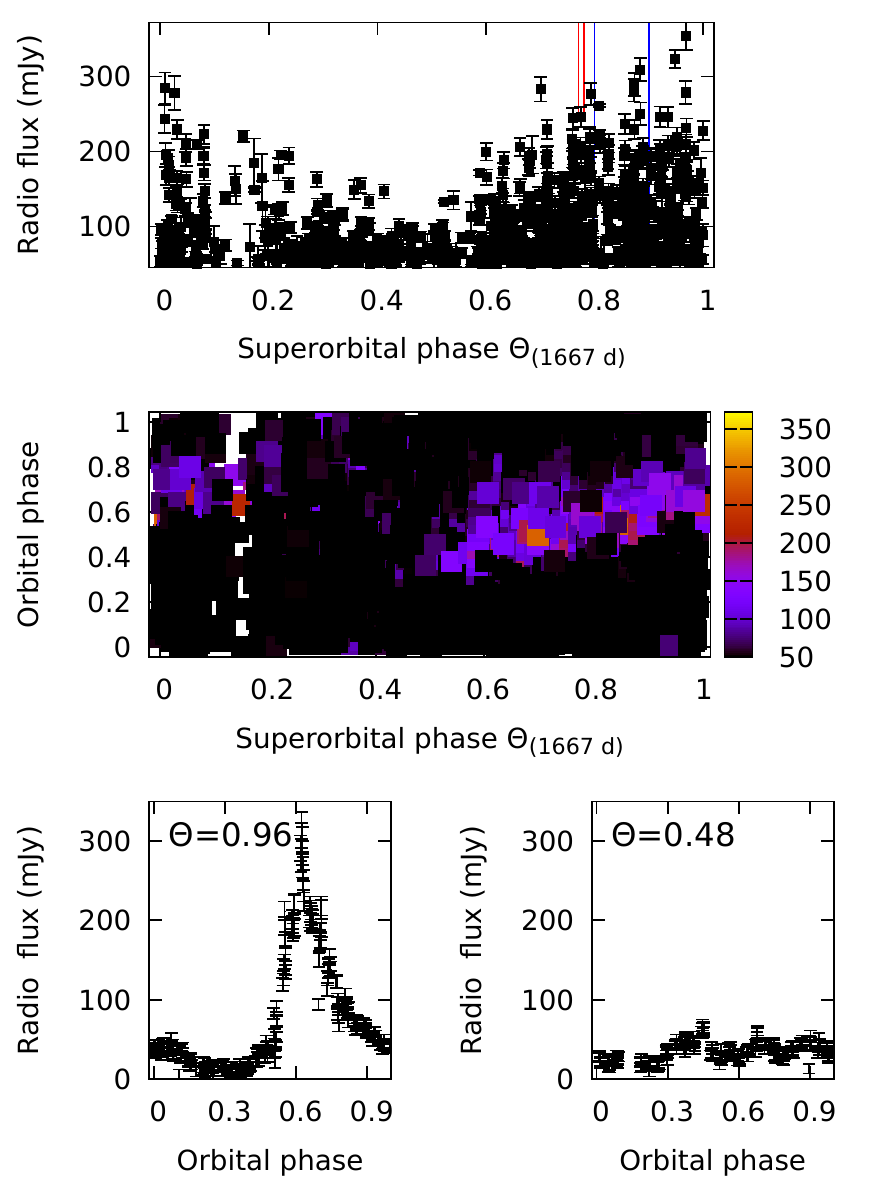}\\
\caption{Long-term modulation in  \lsp. 
Top: 37 yr of radio data 
 \citep{massitorricelli16}
folded with                 
 $ \Theta = {{t-t_0}\over 1667} - {\rm int} ({{t-t_0}\over 1667})$ 
where, $t_0=43366.275$ MJD. 
The two red  lines correspond to the phase interval $\Theta=0.77-0.78$ of  our campaign.
The two blue lines correspond to the interval $\Theta=0.8-0.9$, where simultaneous archival X-ray and $\gamma$-ray data are available (see Sect.~2.3).
Center: Orbital shift of the  radio outburst peak; orbital phase is equal to
 ${{t-t_0}\over P_{\rm orbit}} - {\rm int} ({{t-t_0}\over  P_{\rm orbit} })$.
Bottom: Radio light curves  of \lsp{} at $\Theta_{\rm maximum}$ and $\Theta_{\rm minimum}$.
\label{fig:F1}
}
\end{figure}

The two periodicities, $P_{\rm orbit}$ and $P_{\rm precession}$,
are the two dominant spectral features 
 in the radio power spectrum  
 \citep[37 yrs of observations see][]{massitorricelli16} 
whereas their beat frequencies $P_{\rm average}$ and  $P_{\rm long}$ dominate
the observed emission. 
\citet{ray97} were the first to observe that the radio outbursts in \lsi 
occur with a period of  26.7~days and not with the orbital one. 
The radio outburst periodicity is equal to
 $P_{\rm average}=2/(\nu_{\rm orbit} + \nu_{\rm precession})$ \citep[e.g.,][]{jaron13} 
which 
results in an  orbital phase shift (center panel of Fig.~\ref{fig:F1}) of the radio outburst 
when it is  plotted versus the orbital phase
\citep{massijaron13, massitorricelli14}.
Moreover,  the amplitude of the outburst changes periodically with a 
long-term periodic modulation $P_{\rm long} = 1667 \pm 8$\,days \citep{gregory02} 
    (top panel of Fig.~\ref{fig:F1})  
equal to $1/(\nu_{\rm orbit} - \nu_{\rm precession})$ \citep{massijaron13}. 
In the precessing microquasar scenario the explanation of the beat between 
$P_{\rm orbit}$ and $P_{\rm precession}$ is straightforward: 
The maximum of the long-term modulation is reproduced when the jet gets refilled along the orbit 
($P_{\rm orbit}$) when forming its minimum angle with respect to the line of sight, 
i.e., maximum Doppler boosting. 
Because of precession ($P_{\rm precession}$), one orbital cycle ($P_{\rm orbit}$) 
later the angle is slightly different and the Doppler boosting is reduced. 
Only after $P_{\rm long}$ the system comes back to the conditions that give again 
maximum flux density during the radio outburst
\citep[e.g.,][]{massitorricelli14}.

\begin{figure}
\begin{center}
\includegraphics{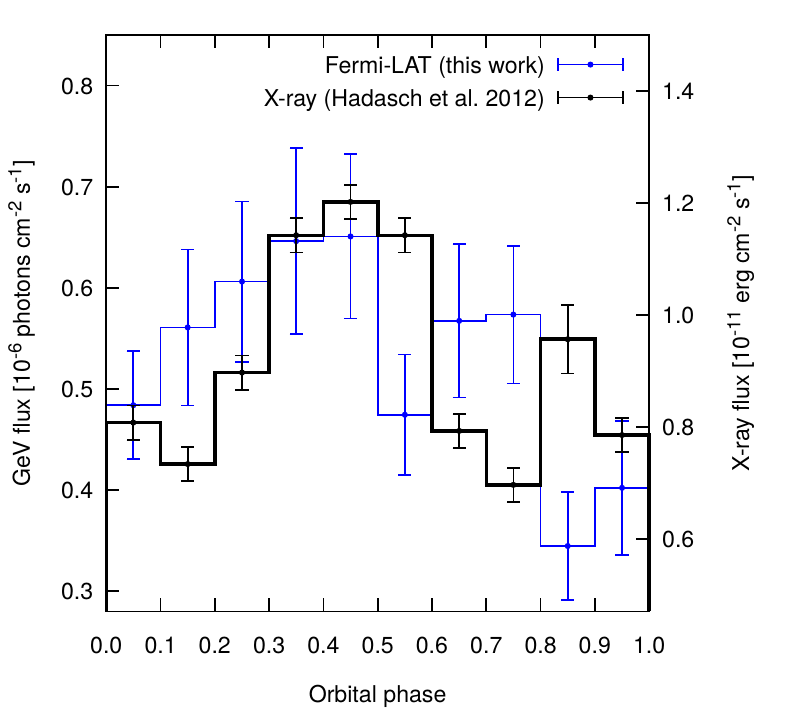}
\caption{
Simultaneous  RXTE-PCA X-ray (black) \citep{hadasch12} and \fermilat GeV $\gamma$-ray (blue)  data  at
$\Theta=0.8-0.9$. The \textit{Fermi}-LAT data are the result of fitting 
photon data from the energy range 0.1-300~GeV 
with a log parabola spectrum (Sect.~2.3).}
\label{fig:hist}
\end{center}
\end{figure}

\citet{massikaufman09} using 6.7~years of Green Bank Interferometer (GBI) radio data at two frequencies
analysed  the spectral index,   function of the ratio of the two flux densities.
They demonstrated that twice along the orbit, during the large outburst 
and closer to periastron when there is only a low radio emission level, 
the radio emission attains the same spectral characteristics, 
resulting in a double-peaked shape of the orbital modulation of the radio spectral index. 
Two peaks are also evident at high energy. The folded light curve of 5~years of \fermilat{} data shows two peaks along the orbit \citep{jaron16}. X-ray and $\gamma$-ray data covering the interval between the two blue lines in the upper panel of Fig.~\ref{fig:F1}, i.e., seven orbital cycles, again hint for two peaks (Fig.~\ref{fig:hist}), where the $\gamma$-ray peak at orbital phase 0.6$-$0.8 seems to be in the dip between the two X-ray peaks. \citet{taylor92} were the first to note that if the compact object in \lsi is accreting, then the accretion rate might develop two peaks along the eccentric orbit.
The accretion rate is given by
\beq
\dot M \propto {\rho_{\rm w} \over v_{\rm rel}^3}
\eneq
where $\rho_{\rm w}$ is the density of the stratified wind of the B0 Ve star at the position of the accretor, and $v_{\rm rel}$ is the relative velocity between the wind and the orbiting accretor \citep{Bondi52}. The accretion rate peaks around periastron, where the density is the highest. At high eccentricities, however, there exists a second accretion rate maximum along the orbit when the compact object is traveling slower and thus the decrease of its velocity relative to the wind particles compensates for the drop in the wind density.
  Flares at periastron and apastron, as observed in Circinus X-1, were explained in a similar way \citep{tudose08}.
This hypothesis for \lsi was further corroborated and modeled by other groups \citep{martiparedes95, boschramon06}. With an independent approach, \citet{romero07} using a Smoothed Particle Hydrodynamics  code, found two peaks in the accretion rate of \lsp. Each peak in accretion rate should be followed by an ejection of particles forming a jet, radiating radio emission (via synchrotron process) and $\gamma$-rays (via inverse Compton, IC, process) \citep{boschramon06, jaron16}. IC losses are stronger close to periastron, leading to the weaker radio emission, in agreement with the results of 5~years of OVRO radio monitoring  
\citep{jaron16}. The second ejection
occurs at a later orbital phase, when accretor and donor star are more displaced from each other. 
Under this condition relativistic electrons can propagate to a region out of the orbital plane and produce a large radio outburst \citep{boschramon06, jaron16}. From the two-accretion peak model with IC losses the large radio outburst results to be associated with a synchrotron emitting jet one order of magnitude larger than the synchrotron emitting jet produced around periastron \citep{jaron16}. Nevertheless, the radio spectral characteristics of the two jets are the same as proven 
by GBI observations \citep{massikaufman09}. 

Thus existing data support the two-peak accretion model already. However, archival radio, X-ray, and $\gamma$-ray observations are not simultaneous and sparse sampling forces one to fold and average several orbital cycles together. Aimed to determine coincidence or possible offsets between different energy bands in the orbital occurrence of the two outbursts predicted in the accretion scenario, we performed a multiwavelength campaign along one orbit (see Table~\ref{T1}). The details of the campaign are given in Sect.~2, and in Sect.~3 we present our results. The conclusions are given in Sect.~4.

\section{Data}

\subsection{Radio observations: 100-m Effelsberg radio telescope and OVRO}

The observations with the 100-m radio telescope at Effelsberg were performed with the
receivers at 11~cm, 6~cm, and 2.8~cm wavelength (S110, S60, S28). All three systems are located
in the Gregorian focus of the telescope (see Table~\ref{T3} for receiver properties).
As \lsp{} as well as the calibrators are point-like for the telescope, the measurements could be done as cross-scans in azimuth and elevation (with a various number
of subscans) over the source positions. At 11\,cm the scanning was done in the equatorial
system to avoid confusion with nearby sources.
For the data analysis, a Gaussian function was fitted to the antenna temperature measurement of each subscan (representing the convolution of the telescope beam with a delta function). The amplitudes of all subscans were corrected for the pointing deviation and averaged, resulting in a single antenna temperature per scan.
Further corrections were applied for the atmospheric attenuation of the signal and for the gain-elevation effect (loss of sensitivity due to
gravitational deformation of the antenna when being tilted). Finally, the conversion factor
from antenna temperature into flux density (in Jy) was determined by the observations of
NGC\,7027, 3C\,286, 3C\,295 and 3C\,48, 
3C\,286 and DR\,21 \citep{baars77}, this factor was applied to all
observations (on a day-by-day basis). Typical values for
these corrections are given in Table~\ref{T3}.  
Typical rms deviations for calibrators are 2$\%$ and for \lsi of 5$\%$.

\begin{table}
\centering
\caption{Log of the  observations in August/September 2017. Column 1: telescopes and satellites.
Columns 2 and 3:  start and stop MJD time. 
We used SWIFT data up to 76 days before the campaign and up to 55 days after the campaign. 
Column 4:  frequency  or energy bands.}
\begin{tabular}{c c c c  }     
\hline\hline
 &  Start      &  Stop  &Energy/Frequency       \\
 &   [MJD]     &  [MJD] &      \\
\hline
100-m Effelsberg  & 57981.7 &  58008.7 & (2.64, 4.85, 10.45) GHz  \\
radio telescope &  &   &  \\
OVRO & 57981.7& 58008.7 & 15 GHz \\ 
\fermilat & 57978.7 &  58005.2 & 0.1-300 GeV \\
\swift  & 57918.5 &  58050.3  & 0.3$-$10 keV \\
\xmm  & 57984.0 &  57984.3 &0.2-15 keV  \\
\nus  &57980.0  &  57980.6 &3-78 keV  \\
\hline
\end{tabular}
\label{T1}
\end{table}

\begin{table}
\centering
\caption{100-m Effelsberg telescope parameters}
\begin{tabular}{c c c c }     
\hline\hline
RX & 11\,cm & 6\,cm &  2.8\,cm  \\
\hline
center freq. &		2.675 GHz &		4.85 GHz	 &	10.45 GHz \\
bandwidth	 &	10 MHz	 &		500 MHz	 &	300 MHz \\
T$_{\rm sys}$ (zenith)	 &	29.3	 &			25.8	 &		28.9 \\
typical opacity		 & 0.01 &				0.015	 &	0.022 \\
 gain-loss$^*$ & $\leq$ 1.2 \% & $\leq$ 1.5 \% & $\leq$ 2.5 \% \\   
sensitivity &		1.5 K/Jy &			1.5 K/Jy &		1.35 K/Jy \\
\hline
\multicolumn{4}{l}{$^*$ due to gravitational deformation (elevation dependent)}\\
\end{tabular}
\label{T3}
\end{table}

The Owens Valley Radio Observatory (OVRO) 40-Meter Telescope uses off-axis
dual-beam optics and a cryogenic receiver with 2~GHz equivalent noise
bandwidth
centered at 15~GHz. Atmospheric and ground contributions, as well as gain
fluctuations, are removed with the double switching technique
\citep[see references in][]{richards11},
where the observations are conducted in an
ON-ON
fashion so that one of the beams is always pointed onto the source. Until
May 2014 the two beams were rapidly alternated using a Dicke switch. Since
May 2014, when a new pseudo-correlation receiver replaced the old
receiver, a
180~degree phase switch was used.
Relative calibration is obtained with a temperature-stable noise diode to
compensate for gain drifts. The primary flux density calibrator is
3C\,286, with
an assumed value of 3.44~Jy \citep{baars77}, 
DR\,21 is used as
secondary calibrator source with an assumed value of 19.04~Jy based on the first four years of OVRO monitoring.
 Details of the observation and data reduction 
schemes are given in  \citet{richards11}.

\begin{figure*}
  \includegraphics[scale=0.96]{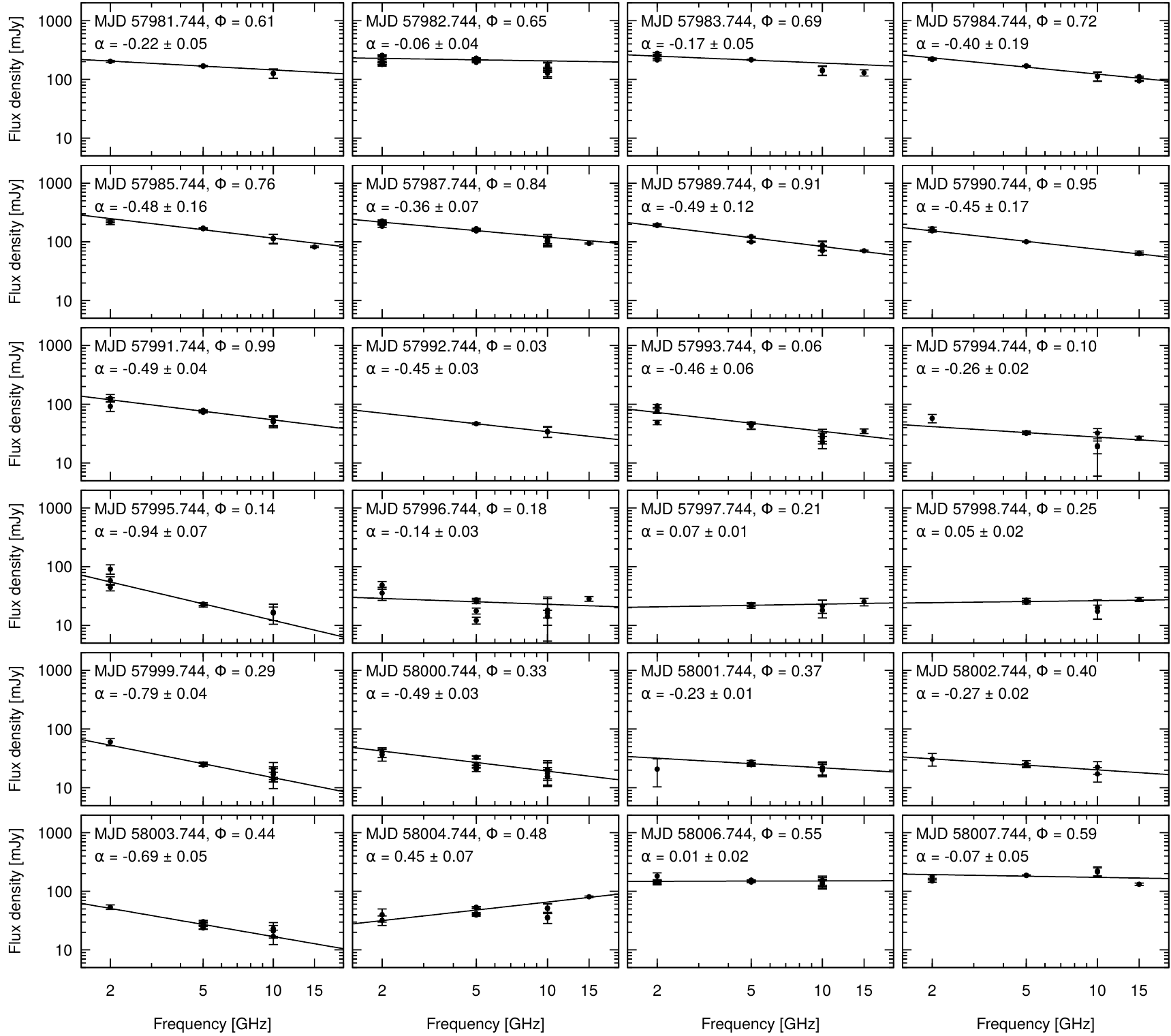}
  \caption{Radio spectra of \lsp{}, as explained in Sect.~2.1.}
  \label{fig:spectra}
\end{figure*}

In order to compute the spectral index\ $\alpha$ from the four radio light curves (three from Effelsberg, one from OVRO), we divided these data into one day time bins and performed an error-weighted least-squares fit of a straight line of the form: $\log S = \alpha\log\nu + b$, to all of the data in $\log \nu$-$\log S$-space within each time bin. We define the reference time for the spectral index value as the center of the time-bin. The uncertainties in $\alpha$ result from formal error propagation. 
Figure~\ref{fig:spectra} shows the radio data and the fitted spectrum for each time bin.

\subsection{X-ray data: \swiftp, \nus and \xmm}

Publicly available \swift data on \lsi have been taken between MJD 57918
and MJD 58050. The data were reprocessed with \texttt{xrtpipeline v.0.13.4} as suggested by the \swift 
team\footnote{See e.g., \swift User's Guide  \url{https://swift.gsfc.nasa.gov/analysis/xrtswguidev1 2.pdf }}.
Spectra were extracted with \texttt{xselect} from a  $36''$-radius circle
around \lsi for source counts and an annulus also centered at the source position with inner
 (outer) radii of $60''$ ($300''$) for background counts. 
The analysis of the  TOO \xmm observation of \lsi (taken on July 19, 2017) was
performed with the \xmm \texttt{Science Analysis software \footnote{https://www.cosmos.esa.int/web/xmm-newton/what-is-sas} v.15.0.0}. Known hot
pixels and electronic noise were removed, and data were filtered to exclude
soft proton flares episodes. The total exposure is $\sim 25$~ksec. The
spectrum and lightcurve (100~sec time bins) were extracted from a $40''$ radius circle centered at the position of
\lsi and the background was extracted from a nearby source-free region of $80''$ radius. The RMFs and ARFs were extracted using the \texttt{RMFGEN} and \texttt{ARFGEN} tools, respectively. 
\nus TOO observation was performed on \lsp, on July, 14th, 2017. The raw data
were processed with standard pipeline processing (HEASOFT v.6.22 with the
NuSTAR subpackage v.1.8.0). We applied strict criteria for the exclusion of
data taken in the South Atlantic Anomaly (SAA) and in the ``tentacle''-like
region of higher background activity near part of the SAA. Level-two data
products were produced with the \texttt{nupipeline} tool with the flags
\texttt{SAAMODE=STRICT} and \texttt{TENTACLE=yes}. The lightcurves (100~sec
time bins) were
extracted for a point source with the \texttt{nuproducts} routine. The
corresponding background flux was derived from a ring-like (inner/outer radii
of $80''$/$196.8''$) region surrounding the source. The analysis was
performed in the energy range of $3-78$~keV. 

We model the data with an absorbed power law model (\texttt{phabs*po} Xspec model) with a free hydrogen column density ($N_H$), slope and model normalization. This model  describes  all considered data sets well, with the reduced $\chi^2\sim 1$. Similar to \citet{chernyakova17} we did not find any clear $N_H$ dependence on the orbital phase. We also performed the search for an additional black body component in high-statistic \xmm TOO observation. We did not find any firm evidence for this component, although please note that the best-fit with black body temperature of $T_{bb}\approx 2$~keV formally improves the fit by $\Delta\chi^2\approx 6$, which corresponds to $ 2\sigma$ detection significance.  

We performed  three fits of the X-ray light curve: 
a fit with a constant,  a constant plus one Gaussian, and two Gaussians, 
that result in a reduced $\chi^2$ of 8.0,  6.1 and  4.2 respectively. 
To decide whether the model 3 
fits the data better than the other models, the  difference-$\chi^2$  test was performed between models 
1 and 3, and between models 
2 and 3. In both cases we found a $p$ less than 10$^{-5}$, where the difference is significant for $p < 0.05$.
The model with the two Gaussians (Table 3 and red line in Fig. 5 A) is:
\begin{eqnarray}
    f(\Phi) & = &  A\rm{e}^{-\frac{1}{2}\frac{\left(\Phi - \Phi_{\rm I}\right)^2}{\sigma_{\rm I}^2}}
     + B\rm{e}^{-\frac{1}{2}\frac{\left(\Phi - \Phi_{\rm II}\right)^2}{\sigma_{\rm II}^2}}
\end{eqnarray}
As discussed  in Sect. 4 (items 1$-$7)  a  physical model of X-ray emission
should take into account   the
two-peak accretion model,  viscousity timescales, the accretion flow emission
and likely a weak jet emission. This will be done in future investigations.

\subsection{\fermilat $\gamma$-ray data}
For the GeV data analysis we use Pass 8 \fermilat{} photon data downloaded from the \fermilat{} 
data 
server\footnote{\url{https://fermi.gsfc.nasa.gov/ssc/data/access/}}. 
We analyze the data using version v11r5p3 of the 
\textit{Fermi} ScienceTools\footnote{\url{https://fermi.gsfc.nasa.gov/ssc/data/analysis/software/}}.
 We include all photon data from 10$^\circ$ around the position of \lsp. We fit \lsi with a log-parabola of the form:
${{\rm d}N}/{{\rm d}E} = N_0\left({E}/{E_{\rm b}}\right)^{-(\alpha + \beta\log(E/E_{\rm b}))}$,
with normalization factor $N_0$, $\alpha$, and $\beta$ left free for the fit, 
and the scale parameter $E_{\rm b}$ fixed to its catalog value. 
All parameters of sources from the 3FGL catalog 
within $10^\circ$ around \lsi were left free for the fit as well. 
Sources between $10^\circ$ and $20^\circ$ were 
included in the analysis with their parameters fixed to their catalog values. 
In order to limit contamination from Earth limb 
photons we excluded time epochs when the target source was observed at a zenith angle greater than $90^\circ$. 
The Galactic diffuse emission was modeled with \textit{gll\_iem\_v06.fits} and the template \textit{iso\_P8R2\_SOURCE\_V6\_v06.txt}. 
We divide the time interval of Table~\ref{T1}  into 12 hours time bins, 
and perform an unbinned likelihood 
analysis
\footnote{\url{https://fermi.gsfc.nasa.gov/ssc/data/analysis/scitools/likelihood_tutorial.html}} 
in each one of these time bins. 
\begin{figure}
  \begin{center}
    \includegraphics{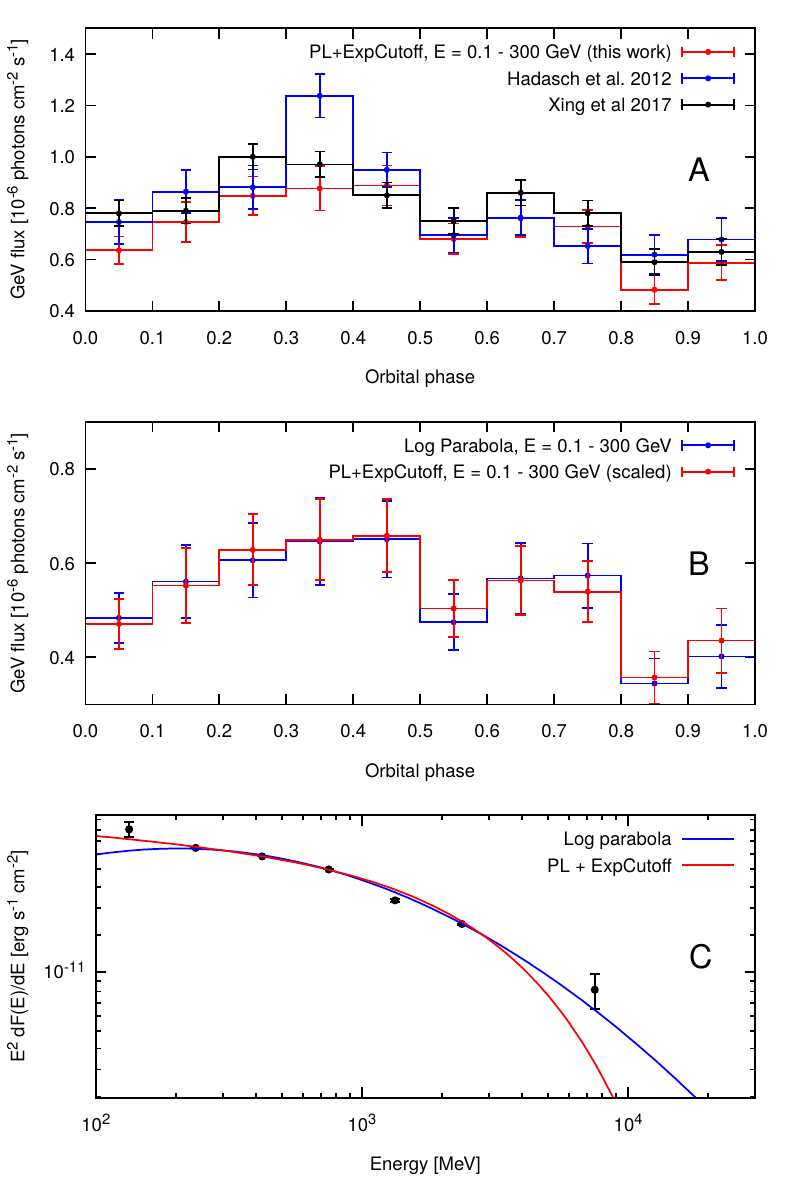}
    \caption{
      A: Comparison of results of different groups using the same spectral fitting model and the same interval of \fermilat{} data  ($\Theta = 0.8-0.9$).
      B: Comparison of the same data processing but two different spectral fitting models. The Log Parabola is the model used for \lsi in the \fermilat{} catalogs. The red curve is scaled by 1/1.35.
      C: Power spectrum and fits (see Sect.~2.3).
      \label{fig:compare}
    }
  \end{center}
\end{figure}

We compared our results with ones that were previously published by \citet{hadasch12} and \citet{xing17}. Figure~\ref{fig:compare}~A  shows three curves obtained by us, \citet{hadasch12}, and \citet{xing17}, all three using  the same data set (MJD 54683-54863) and model  (power law with exponential cutoff). As one can see, in the orbital phase bin  0.3-0.4, the value of Fig.~11  in \citet{hadasch12} is at about 2$\sigma$ displaced from the values of the other two curves \citep[ours and from][]{xing17}, otherwise the three curves all agree  within the errors.
Since 2012 the recommended model for \lsi in the \fermilat catalog is the log-parabola \footnote{Table~4 in \url{https://fermi.gsfc.nasa.gov/ssc/data/access/lat/4yr_catalog/}}. We therefore checked whether two different models give rise to significant differences, and found that the resulting orbital lightcurves have the same shape, but differ by a factor 1.35, see Fig.~\ref{fig:compare}~B. The PL plus exponential cut-off better describes data around 100 MeV, as shown in the  Fig.~\ref{fig:compare}~C. The global fit is however better for the Log parabola model (with $\chi^2$ of 18 with respect to that of 21 for the PL+ExpCutoff).


\begin{table}
    \centering
       \caption{Fit parameters of Eq. 2} 
    \begin{tabular}{lr}
         \hline
         \hline
         Parameter & Value \\
         \hline
         $A$        & $(2.0 \pm 0.2)\times 10^{-11}$\\
         $\Phi_{\rm I}$   & $0.35 \pm 0.01$\\
         $\sigma_{\rm I}$ & $0.11 \pm 0.02$\\
         $B$        & $(1.4 \pm  0.1)\times 10^{-11}$\\
         $\Phi_{\rm II} $   & $0.84 \pm 0.01$\\
         $\sigma_{\rm II} $ & $0.14 \pm 0.02$\\
         \hline
         \hline
    \end{tabular}
    \label{tab:fit}
\end{table}

\begin{table}
\centering
\caption{ Accretion-ejection in \lsp.} 
\begin{tabular}{c c c}     
\hline\hline
 &  Orbital    phase & Orbital phase       \\
 & First event (I) & Second event (II)   \\
\hline
X-ray dip & 0.5-0.7 &  0.0-0.2  \\
$\gamma$-ray onset & 0.5-0.6 & 0.0-0.15  \\
radio  onset  & 0.5-0.7  &  $\sim$0.15  \\
$\alpha \geq$ 0 & 0.5-0.55 &  0.2  \\
\hline
\end{tabular}
\label{T2}
\end{table}

\section{RESULTS}

\begin{figure}
\includegraphics{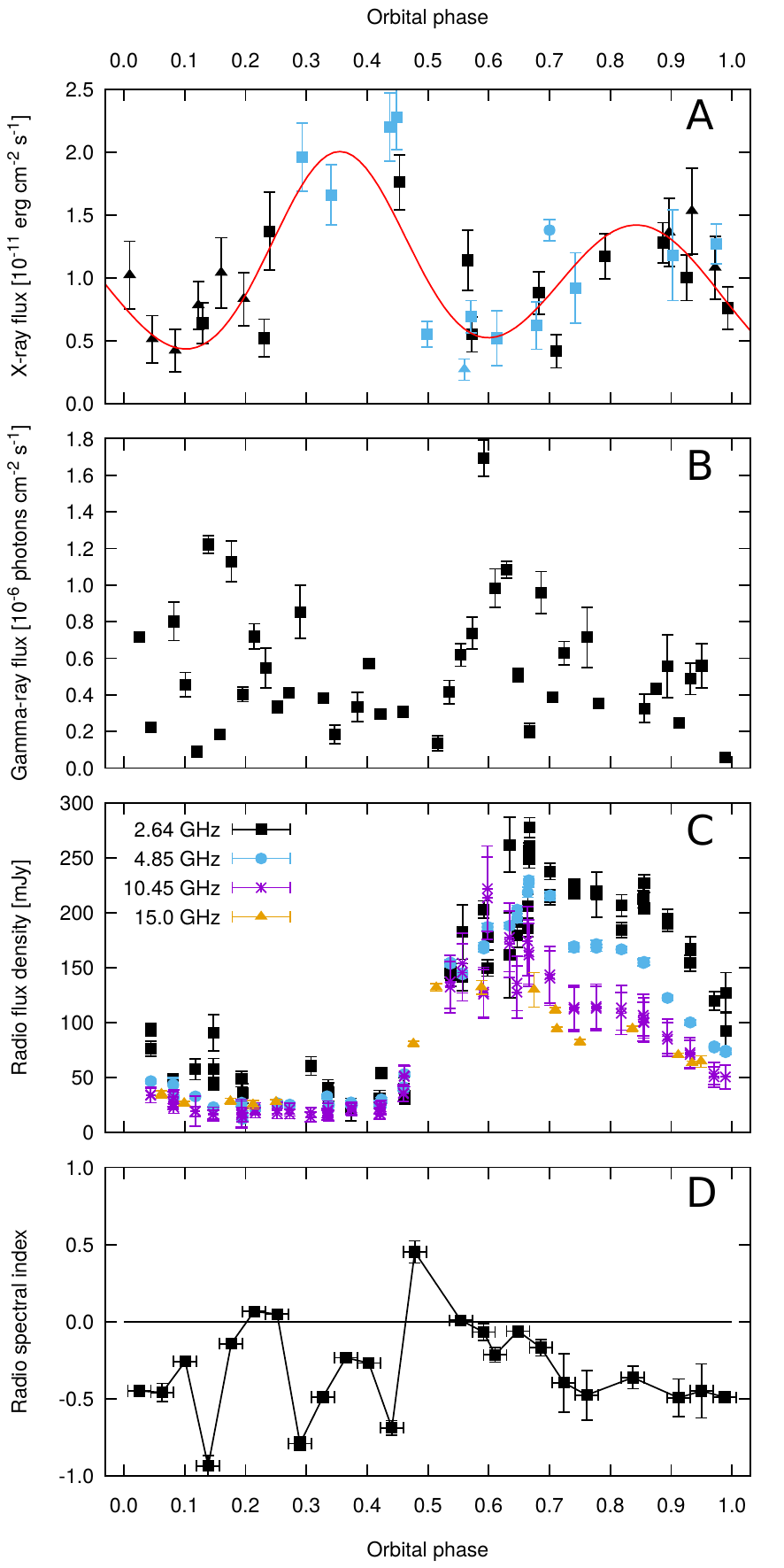}
\caption{Multiwavelength campaign on \lsp{}. Data  vs orbital phase:
A) X-ray data: \swiftp, \xmm and \nus  campaign data 
(blue colour: squares, circles, triangles). Swift data up to 76 days before the campaign use black squares
and up to 55 days after the campaign use black triangles 
(the red line profile shows a simple model with two Gaussians,  see discussion in Sect. 2.2)
B) \fermilat{} data.
C) Radio data: 100-m Effelsberg telescope (2.64, 4.85 and 10.45 GHz) and OVRO (15.0 GHz).
D) Radio spectral index.}
\label{fig:lc}
\end{figure}

In Fig.~\ref{fig:lc} we present radio, X-ray and $\gamma$-ray results.
Figure~\ref{fig:lc}~A shows two X-ray peaks.  One X-ray peak  is at $\Phi \sim 0.35$
with a dip at $\Phi = 0.5-0.7$. 
The X-ray emission rises again, peaks at $\Phi \sim 0.85$   and drops again
at $\Phi = 0.0-0.2$.  
The  X-ray data of Fig.~\ref{fig:lc}~A confirm  the prediction of the
  accretion theory in \lsi{}  
\citep{taylor92, martiparedes95, boschramon06, romero07}.
In fact the accretion flow emits in X-ray \citep[e.g.,][]{yang15a,yang15b},
therefore the predicted  increase in the accretion rate
implies an increase in the X-ray emission twice along the orbit.

Figure~\ref{fig:lc}~B shows two $\gamma$-ray peaks rising
 above a scattered level of about $4~10^{-7}$ photons cm$^{-2}$ s$^{-1}$.
This is not the first observation of two $\gamma$-ray peaks along the orbit of \lsp.
In the folded curve of 5 yr of $\gamma$-ray data \citep[left panel of Fig.~7 in][]{jaron16}
two peaks arise above an offset of
$5~10^{-7}$ cm$^{-2}$ s$^{-1}$.
The two peaks were modeled with two ejections of particles emitting $\gamma$-ray emission due to IC process \citep{jaron16}
and the offset explained as a contribution from the accretion flow population \citep[see Sect.~5 in][]{jaron16}.
The new information coming from Fig.~\ref{fig:lc}~B
is about the misalignment between the $\gamma$-ray peaks associated to the ejection of particles with respect to the X-ray peaks. 
As summarized in Table~\ref{T2}, it is during the X-ray dip at $\Phi=0.5-0.7$ that occurs the $\gamma$-ray onset, $\Phi=0.5-0.6$ and
it is during the X-ray dip at $\Phi=0.0-0.2$
that occurs  the $\gamma$-ray onset,  $\Phi=0.0-0.15$.  
The result of the campaign of $\gamma$-ray peaks during the X-ray dips, probes and gives a better understanding of  the misalignment suggested from archived data folding and averaging all together 
seven orbital cycles affected by different Doppler boosting effects (Sect.~1).
 One can compare these results of  archived data  -- taken 9~years before our campaign --  with our present results.
The curve in Fig.~\ref{fig:hist} hints for two X-ray peaks and two $\gamma$-ray peaks,
these two emerging from an offset of $3-5~10^{-7}$ photons cm$^{-2}$ s$^{-1}$.
The first X-ray peak in Fig.~\ref{fig:hist} occurs at $\Phi=0.45\pm0.15$ (campaign  peak at $\Phi=0.35\pm0.01$),
with  the associated $\gamma$-ray peak at $\Phi=0.7\pm0.1$ (campaign  peak at $\Phi \sim 0.6$).
The second  X-ray peak occurs at $\Phi=0.85\pm0.05$ (campaign peak at  $\Phi=0.84\pm0.01$),
with  the associated $\gamma$-ray peak at $\Phi=0.3\pm0.2$ (campaign peak at  $\Phi \sim 0.15$).

Figure~\ref{fig:lc}~C shows a large radio outburst with onset (Table~\ref{T2}) during the X-ray dip at $\Phi=0.5-0.7$.
As discussed in Sect.~1 the accretion theory for \lsi predicts that the two peaks in the accretion rate are  followed by ejections
of electrons experiencing strong energetic losses when  ejected close to periastron passage, because
of IC interactions with  UV stellar photons.
As a result there is  only one large radio outburst,  associated
to  the ejection more displaced from periastron
\citep[e.g.,][]{jaron16}.
The new result of our Fig.~\ref{fig:lc} is the discovery 
that the onset of the radio outburst, as that of the associated $\gamma$-ray peak,
occurs in the dip of the X-ray emission.

\begin{figure}
\begin{center}
\includegraphics[scale=0.9]{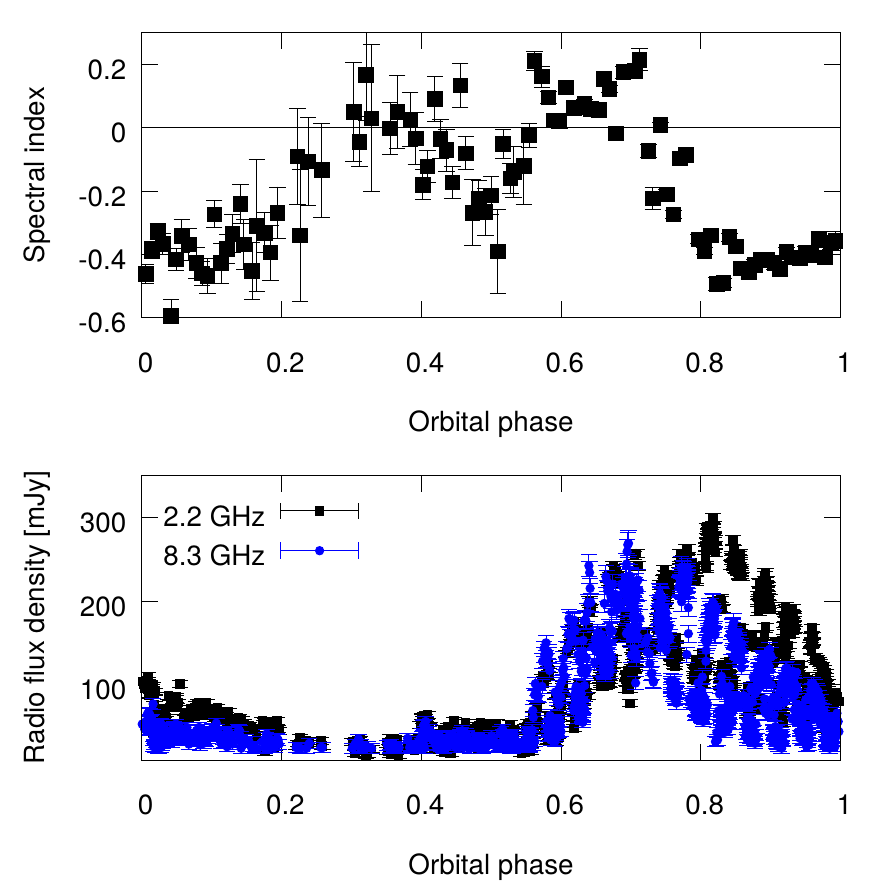}
\caption{
Double-peaked shape of the  spectral index of \lsp. 
Top: Spectral index data (averaged over $\Delta \Phi=0.009$). 
Bottom: GBI data in the interval 50048-50174 MJD.
\label{fig:alpha}
}
\end{center}
\end{figure}

The result of \citet{jaron16} is that electrons ejected closer to periastron, because of their strong energetic losses, produce a short radio jet,
i.e., more than one order of magnitude shorter
than that produced in the second ejection, more displaced from the B0 star, and responsible for the large radio outburst.
Different in their flux density, still the two jets present both the typical spectral characteristics  of jets in microquasars
of a flattish or inverted spectrum.
In Fig.~\ref{fig:alpha}  we show 5 cycles  of \lsi GBI data folded with the orbital phase. 
The spectral index  changes from negative to  positive 
at the onset of the large outburst 
and also around orbital phase 0.3, even if there  only low emission is present.
The campaign allows us to analyse   for the first time the radio spectral
index along one complete orbit at three to four radio frequencies instead of the two frequencies used in the GBI  observations.
The multiwavelength radio spectra show indeed that the radio spectral index becomes $\alpha \geq 0$ at $\Phi\sim 0.2$ and $\Phi=0.5-0.55$.

\begin{figure}
\begin{center}
\includegraphics[scale=0.9]{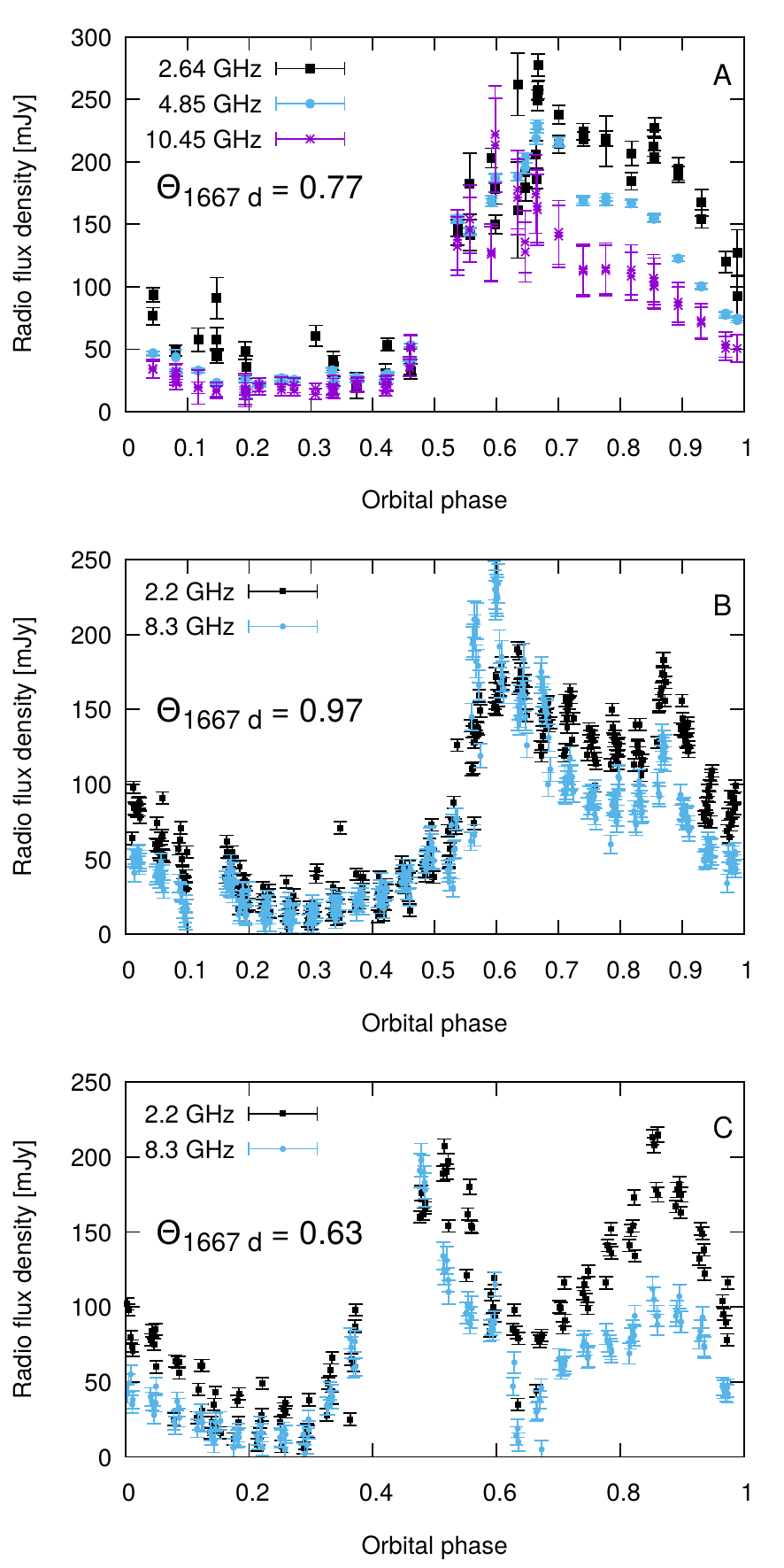}
\caption{
Evolution from the self-absorbed jet to the transient jet in \lsi at three
epochs. A: campaign radio data B: GBI data starting at MJD
49995 C: GBI data starting at MJD 51080.
\label{fig:transient}
}
\end{center}
\end{figure}

In  microquasars after the flat spectrum phase a shock-related
 transient, giving an optically thin outburst (Sect.~1),  might develop. Figure~\ref{fig:lc}~D shows that the 
self-absorbed jet, i.e., $\alpha \geq 0$, clearly persists until $\Phi\simeq 0.55$, after that the 
   radio spectral index  oscillates and  stabilizes to $\alpha \simeq - 0.5$ for an optically thin outburst.
Figure~\ref{fig:transient} compares the campaign with other two epochs: in the campaign (top panel of Fig.~\ref{fig:transient}) the optically thin outburst forms
just a bump during the decay of the radio light curve, in 
Figs~\ref{fig:transient}~B and~C the onset of the optically thin outburst, and likely onset of the shock,
is at a later phase during  the decay of  the flat-inverted spectrum outburst.
A systematic study of the system \lsi will bring a deeper understanding of the physical conditions triggering the transient.

\section{CONCLUSIONS AND DISCUSSION}

Theory predicts two events of accretion$-$ejection along the orbit of \lsp.
From archival data, showing the two predicted events, one can only draw broad conclusions because the
observations at the various energy bands are not simultaneous.
We performed a multiwavelength campaign to  establish  their relative orbital
occurrence. Our results are:
\begin{enumerate}
\item{
  X-ray emission rises  twice along the orbit, as predicted by the model. After reaching the maximum the emission drops and two dips are noticeable along the orbit.
  }
\item{
  It is during the two X-ray dips that $\gamma$-ray emission increases. 
}
\item{
  The radio spectral index $\alpha$ becomes greater or equal to zero twice along the orbit, revealing that a self-absorbed jet is generated twice. The large radio outburst is displaced from periastron passage and happens during the X-ray dip.
}
\end{enumerate}

X-ray dips associated to  radio emission peaks were  observed 
 in the stellar mass
black hole system \grs{} \citep{mirabelrodriguez99, kleinwolt02} and
 explained \citep[][and references therein]{kleinwolt02} as disappearance of matter from the accretion flow (the X-ray dip),
part of it passing through the event horizon and part of it ejected into a radio emitting jet (the radio peak).
\citet{marscher02} interprete in a similar way four radio ejections appearing after four X-ray dips in the AGN 3C 120, i.e., as observational evidence for their accretion-disk origin.
It is worth of note that in our multiwavelength campaign, we also see a $\gamma$-ray peak together with the radio peak.

Although jets are ubiquitous in the Universe,
common to a number of different accreting systems, the mechanism of jet production
and the details of the accretion-ejection processes are still poorly constrained.
\lsi could become  the laboratory
for  understanding
the accretion-ejection coupling:
periodic, detected at all wavelengths from the radio band to very high-energy $\gamma$-rays
and
a  jet  well traceable
with  radio interferometers.
\\
Several  investigations will be possible: 
\\
\\
1. By means of  X-ray observations of \lsp, it becomes
possible for  the theoretical models to  include and estimate viscous timescales.
The accretion models, based   on  Eq. 1,  predict  the accretion peaks only.
From the delay between predicted peaks in accretion rate and X-ray peaks from the accretion flow, 
a quantitative analysis of the viscous  processes will be possible.
\\
\\
2.  By studing  the relationship between 
 X-ray dip and
radio onset 
 in \lsi it will be possible to study the coupling between accretion and ejection.
\\
\\
3. Optical depth effects  along the jet, as done in Cygnus X-1 \citep{tetarenko19},
  can be well analysed in the flat-inverted phase of the outburst.
\\
\\
4. A better understanding of  shock propagation along the  jet will
 be possible analysing the delays between self-absorbed  jet and transient jet.
\\
\\
5. The possible relationship between  transient jet and   emission at very high energy can be studied.
In fact TeV emission is thought to be produced by shocks \citep[e.g.,][]{williamson19} 
and in microquasars shocks are associated with the transient jet
\citep{fender04}.
It is therefore  during a transient jet that TeV emission could be expected.
The system  Cygnus~X-1, 
which with a jet  angle  of 27 degrees
\citep{orosz11} and jet velocity  $0.92c$ \citep{tetarenko19} corresponds to a Doppler boosting 
of about 5,
shows mostly the self-absorbed radio jet \citep{stirling01}. 
Indeed, in a set of several MERLIN observations \citet{fender06} were able at only one epoch to pick 
up the system during a transient.  
Worth of note, in Cygnus X-1 TeV emission was detected (4.0 $\sigma$) \citep{albert07}
 coinciding with a flare seen by RXTE, Swift, and INTEGRAL.
The system \lsi  shows several episodes of a transient jet \citep[e.g.,][]{massi14} 
and emission at TeV energies were in fact detected several times in \lsi \citep[][and references therein]{veritas16}.
The systems Cyg X-1 and \lsi are both high mass X-ray binaries (HMXBs),
suggesting that  the stellar wind and/or the radiation field of the companion star  play an important
role, along with transient jet and Doppler boosting,  in producing the TeV high-energy photons. In leptonic
models these photons may be  produced by inverse Compton  scattering of
the UV stellar radiation or wind X-ray emission off the electrons of the transient jet, whereas in hadronic
models $\gamma$-rays may be produced by the decay of neutral pions associated to
the  interactions of relativistic protons in the jets with  protons of the
stellar wind \citep[][and references therein]{romero17}.
 \\
 \\
\\
\\
6.   The HMXB LSI +61 303 is radio overluminous compared to what is seen in typical black hole LMXBs in their hard states (when the radio spectra are flat) in the same luminosity range.
The X-ray luminosity of \lsp,  
$2.0 \times 10^{33}\leq L_X \leq 1.8 \times 10^{34}$ erg/sec \citep{massiet17},
is nearly at the  boundary between the low/hard X-ray state and
the quiescent state. 
  The quiescent state, 
where, e.g., is located the only other known microquasar accreting from a Be star, MWC 656 \citep{dzib15},
starts  in fact at  about  $L_{\rm X} \sim 10^{33}~ {\rm erg/sec}$ \citep{mcclintockremillard06}.
The models of the low/hard state give a geometrically thin, 
optically thick accretion disk truncated at $R = 100\,r_{\rm g}$ \citep{mcclintockremillard06}.
 Observed oscillations at 2\,Hz by  \citet{rayhartman08} imply in \lsi a larger truncated value,
   $R = 300\,r_{\rm g}$ \citep{massizimmermann10} in agreement with the low luminosity.
Close to the quiescent state,  the emission from the electrons of the
accretion flow  drops to low values and   
an open issue is if jet electrons contribute to the X-ray emission
\citep{yang15a,plotkin17}.
The favourable case of \lsi is that this  jet component can be in fact observed, as discussed in Sect. 1,
 either   with the direct detection in the 
timing analysis of the precession period of the jet or by  
observing related phenomena as the  orbital shift and  the long-term modulation.
Indeed,  
the long-term modulation  of the X-ray emission was observed \citep{chernyakova12}    
and the precession period  detected by timing analysis of X-ray observations \citep{dai16}.

Planning future observations one has to take into account that from the  accretion theory the  sequence of accretion-ejection along the orbit of \lsi
is   a periodic feature, but  on the other hand, in the observer frame
the variable Doppler boosting induced by jet precession changes the observed emission \citep{massitorricelli14}.
Observations  show  that the precession brings the  jet
close to the line of sight \citep[bottom panel of Fig.~10 in ][]{wu18},
inducing maximum Doppler boosting of the  emission
\citep{massitorricelli14}.
The result of beating between precession and periodic ejection  \citep[e.g., Fig.~4 in ][]{massitorricelli14}
is that the radio peak does not only change its amplitude  along the 1667~d period
but also  shifts in orbital phase (center panel of Fig.~\ref{fig:F1}).
As shown in the bottom panel of Fig.~\ref{fig:F1},  around the minimum of the 1667~d modulation
there is only  low broad radio emission formed by  several small peaks.
Therefore, future observations focused to the study of the accretion-ejection coupling should avoid
the minimum of this long-term modulation and observe towards the maximum, as we did in our campaign.

\section*{Acknowledgements}

FJ thanks Walter Alef and Helge Rottmann  for providing computer power at the MPIfR high performance computer cluster.
  We  thank Eduardo Ros  for reading the manuscript and providing helpful comments.
 This research has made use of  
  observations with the 100-m telescope of the MPIfR (Max-Planck-Institut f\"ur Radioastronomie) at Effelsberg,
 \fermilat data obtained from
the High Energy Astrophysics Science Archive Research Center
(HEASARC), provided by National Aeronautics and Space Administration (NASA) Goddard Space Flight Center,
 \xmm data, an ESA science mission with instruments and contributions directly funded by ESA Member States and NASA,
 OVRO 40-m monitoring program 
supported in part by NASA
grants NNX08AW31G, NNX11A043G, and NNX14AQ89G and NSF grants
AST-0808050 and
AST-1109911. The authors acknowledge support by the state of Baden-W\"urttemberg through bwHPC. This work was supported in part by DFG through the grant MA 7807/2-1.

\noindent
{\bf Data availability}\\
The data underlying this article will be shared on reasonable request to the 
corresponding author.

%
\bsp	
\label{lastpage}
\end{document}